%% file: main.tex
\documentclass[conference]{IEEEtran}
\IEEEoverridecommandlockouts
\usepackage{cite}
\usepackage{amsmath,amssymb,amsfonts}
\usepackage{algorithmic}
\usepackage{graphicx}
\usepackage{textcomp}
\usepackage{xcolor}

\usepackage{cite}
\usepackage{amsmath,amssymb,amsfonts}
\usepackage{algorithmic}
\usepackage{graphicx}
\usepackage{textcomp}
\usepackage{xcolor}
\usepackage{float}
\usepackage{svg}
\usepackage{soul}
\usepackage{booktabs}
\usepackage{todonotes}
\usepackage[colorlinks=true, linkcolor=blue, citecolor=blue, filecolor=blue, urlcolor=blue]{hyperref}
\usepackage{array}

\usepackage{algorithm}
\makeatletter
\newcommand{\removelatexerror}{\let\@latex@error\@gobble}
\makeatother

\usepackage{booktabs}

\usepackage{glossaries}
\setacronymstyle{long-short}

\usepackage{pgfplots}
\usepackage{subcaption}

\def\BibTeX{{\rm B\kern-.05em{\sc i\kern-.025em b}\kern-.08em
    T\kern-.1667em\lower.7ex\hbox{E}\kern-.125emX}}
\begin{document}

\title{SSC-UNet: UNet with Self-Supervised Contrastive Learning for Phonocardiography Noise Reduction\\

\thanks{
\IEEEauthorrefmark{3} Corresponding author.\\
This work was supported by the Taighde Éireann – Research Ireland under Grant number 22/PATH-S/10763.}
}


\author{
Lizy Abraham\IEEEauthorrefmark{1}\IEEEauthorrefmark{3}, 
Siobhan Coughlan\IEEEauthorrefmark{2},
Kritika Rajain\IEEEauthorrefmark{1}, 
Changhong Li\IEEEauthorrefmark{1}, 
Saji Philip\IEEEauthorrefmark{3}, 
Adam James\IEEEauthorrefmark{2} \\

\IEEEauthorrefmark{1}Walton Institute, South East Technological University, Waterford, Ireland \\
\IEEEauthorrefmark{2}Children’s Health Ireland Hospital, Dublin, Ireland\\
\IEEEauthorrefmark{3}Tiruvalla Medical Mission Hospital, Kerala, India \\
Email: lizy.abraham@waltoninstitute.ie, siobhan.coughlan@childrenshealthireland.ie, kritika.rajain@waltoninstitute.ie,\\ changhong.li@waltoninstitute.ie, sajitfc@gmail.com, adam.james@childrenshealthireland.ie
}

\maketitle
\begin{abstract}
Congenital Heart Disease (CHD) remains a significant global health concern affecting approximately 1\% of births worldwide.
%
Phonocardiography has emerged as a supplementary tool to diagnose CHD cost-effectively.
%
%
However, the performance of these diagnostic models highly depends on the quality of the phonocardiography, thus, noise reduction is particularly critical.
%
Supervised UNet effectively improves noise reduction capabilities, but limited clean data hinders its application.
The complex time-frequency characteristics of phonocardiography further complicate finding the balance between effectively removing noise and preserving pathological features.
In this study, we proposed a self-supervised phonocardiography noise reduction model based on Noise2Noise to enable training without clean data.
%
Augmentation and contrastive learning are applied to enhance its performance. We obtained an average SNR of 12.98 dB after filtering under 10~dB of hospital noise. Classification sensitivity after filtering was improved from 27\% to 88\%, indicating its promising pathological feature retention capabilities in practical noisy environments.
%
%
\end{abstract}

    \input{myglossaries}
    \input{sec1_introduction}
    \input{sec2_related_works}
    \input{sec3_database_and_method}
    \input{sec4_result_and_discussion}
    \input{sec5_conclusion}
    \input{sec6_ack}

    \bibliographystyle{IEEEtran}
    \bibliography{Ref,IEEEabrv}
\end{document}

%% file: myglossaries.tex
\newacronym{svm}{SVM}{Support Vector Machine}
\newacronym{chd}{CHD}{Congenital Heart Disease}
\newacronym{pcg}{PCG}{Phonocardiography}
\newacronym{snr}{$SNR$}{Signal-to-Noise Ratio}
\newacronym{mse}{$MSE$}{Mean Squared Error}
\newacronym{psnr}{$PSNR$}{Peak Signal-to-Noise Ratio}
\newacronym{cc}{$CC$}{Correlation Coefficient}
\newacronym{ppv}{$PPV$}{Positive Predictive Value}
\newacronym{acc}{$Acc$}{Accuracy}
\newacronym{se}{$S_e$}{Sensitivity}
\newacronym{sp}{$S_p$}{Specificity}
\newacronym{f1}{$f_1$}{F1 Score}
\newacronym{auc}{$AUC$}{Area Under the Curve}
\newacronym{ecg}{ECG}{Electrocardiogram}
\newacronym{pca}{PCA}{Principal Component Analysis}
\newacronym{fft}{FFT}{Fast Fourier Transform}
\newacronym{dft}{DFT}{Discrete Fourier Transform}
\newacronym{stft}{STFT}{Short-Time Fourier Transform}
\newacronym{mfcc}{MFCC}{Mel-Frequency Cepstral Coefficients}
\newacronym{cnn}{CNN}{Convolutional Neural Network}
\newacronym{rnn}{RNN}{Recurrent Neural Network}
\newacronym{lstm}{LSTM}{Long Short-Term Memory}
\newacronym{gru}{GRU}{Gated Recurrent Unit}
\newacronym{dnn}{DNN}{Deep Neural Network}
\newacronym{ann}{ANN}{Artificial Neural Network}
\newacronym{relu}{ReLU}{Rectified Linear Unit}
\newacronym{sgd}{SGD}{Stochastic Gradient Descent}

\newacronym{coa}{COA}{Coarctation of the Aorta}
\newacronym{tof}{TOF}{Tetralogy of Fallot}
\newacronym{pda}{PDA}{Patent Ductus Arteriosus}
\newacronym{asd}{ASD}{Atrial Septal Defect}
\newacronym{tr}{TR}{Tricuspid Regurgitation}
\newacronym{pr}{PR}{Pulmonary Regurgitation}
\newacronym{ps}{PS}{Pulmonary Stenosis}
\newacronym{mvp}{MVP}{Mitral Valve Prolapse}
\newacronym{mr}{MR}{Mitral Regurgitation}
\newacronym{ar}{AR}{Aortic Regurgitation}
\newacronym{vsd}{VSD}{Ventricular Septal Defect}

\newacronym{s1}{$S_1$}{First heart sound}
\newacronym{s2}{$S_2$}{Second heart sound}
\newacronym{s3}{$S_3$}{Third heart sound}
\newacronym{s4}{$S_4$}{Fourth heart sound}

\newacronym{emd}{EMD}{Empirical Mode Decomposition}
\newacronym{dwt}{DWT}{Discrete Wavelet Transform}
\newacronym{svd}{SVD}{Singular Value Decomposition}
\newacronym{vmd}{VMD}{Variational Mode Decomposition}

\newacronym{hmm}{HMM}{Hidden Markov Model}
\newacronym{hsmm}{HSMM}{Hidden Semi-Markov Model}

\newacronym{mlp}{MLP}{Multi-Layer Perceptron}
\newacronym{bilstm}{Bi-LSTM}{Bidirectional Long Short-Term Memory}

\newacronym{lime}{LIME}{Local Interpretable Model-Agnostic Explanations}
\newacronym{shap}{SHAP}{Shapley Additive Explanations}

\newacronym{knn}{KNN}{K-Nearest Neighbors}

\newacronym{pfo}{PFO}{Patent Foramen Ovale}
\newacronym{ms}{MS}{Mitral Stenosis}

\newacronym{nmf}{NMF}{Nonnegative Matrix Factorization}
\newacronym{acrc}{ACRC}{Adaptive Contour Representation Computation}

\newacronym{ssr-ht}{SSR-HT}{Shifted-Symmetrized-Regularized Hard-Thresholding}
\newacronym{adaogs}{AdaOGS}{Adaptive Overlapping Group Sparse}
\newacronym{nkf}{NKF}{Nonlinear Kernel Function}

\newacronym{prd}{PRD}{Percentage Root Mean Square Difference}
\newacronym{ai}{AI}{Artificial Intelligence}

\newacronym{mfsc}{MFSC}{Mel-Frequency Spectral Coefficients}   

\newacronym{rmse}{$RMSE$}{Root Mean Squared Error}

\newacronym{wt}{WT}{Wavelet Transform}
\newacronym{crnn}{CRNN}{Convolutional Recurrent Neural Network}

\newacronym{ewt}{EWT}{Empirical Wavelet Transform}
\newacronym{dt}{DT}{Decision Tree}
\newacronym{cvds}{CVDs}{Cardiovascular diseases}
\newacronym{dl}{DL}{Deep Learning}

\newacronym{cwt}{CWT}{Continuous Wavelet Transform}
\newacronym{tdnn}{TDNN}{Time Delay Neural Networks}

\newacronym{ml}{ML}{Machine Learning}

\newacronym{imfs}{IMFs}{Intrinsic Mode Functions}

%% file: sec1_introduction.tex
\section{Introduction}
%
Congenital Heart Disease (CHD) is a malformation caused by abnormal heart development in utero and remains the most common congenital abnormality\cite{sayasathid2012epidemiology, rossano2020congenital}.
%
Early diagnosis of critical CHD reduces mortality compared to late diagnosis \cite{eckersley2016timing}.
Early screening and echocardiography are the main approaches to diagnosing CHD~\cite{mertens2009gold}.
The accessibility of echocardiography examination in developing regions is limited due to equipment cost and the need for specialized personnel. This is particularly challenging for physicians working in low income countries and amongst indigenous populations~\cite{marangou2019echocardiography}.
Heart sounds are generated by the mechanical activity of the heart valves during the cardiac cycle. They provide diagnostically important information regarding the physiological condition of the heart  \cite{liu2016open}. 
Auscultation with a stethoscope is the traditional method of assessing heart sounds. Its digital version, phonocardiography (PCG), is of paramount importance, being a non-invasive and cost-effective way for preliminary screening of CHDs \cite{humayun2020towards}.

The emergence of deep learning techniques has led to a series of PCG diagnostic models, demonstrating remarkable performance. Auscultation with a digital stethoscope, the collection process of PCG, is frequently affected by interference from environmental noise, lung sounds and patient movement \cite{rouis2019optimal, sujadevi2019hybrid}.
The reduction or elimination of these additional sounds is crucial for the subsequent signal processing, as it ensures the accuracy and reliability of further analysis. Therefore, the development of effective PCG noise reduction techniques is imperative.


Several traditional algorithms have been introduced to de- noise PCG. The fundamental components of PCG are primarily concentrated below 800 Hz \cite{chowdhury2020time}. Consequently, low- pass and band-pass filters are widely employed in PCG noise reduction due to their simple yet effective structures. Advanced denoising techniques, such as Empirical Mode Decomposition (EMD) and Discrete Wavelet Transform (DWT)~\cite{ghosh2020evaluation}, are also commonly used. However, given the complex time-frequency characteristics and low-frequency properties of PCG, many of these methods struggle to eliminate noise while preserving the underlying pathological features.

%
Deep learning also provides a solution for fine-grained PCG noise reduction. U-Net mimics decomposition and re- construction of DWT, demonstrating its denoising and feature preservation capabilities \cite{gonzalez2023robust, ali2023end}.
However, models often attempt to restore the signal to a uniform pattern to achieve low loss across different scenarios under the influence of a single loss function, This will lead to blurred outputs, and occasionally sound ‘hallucinations’. Hallucinations are misleading or incorrect outputs generated by an AI model that appear plausible but are actually false or fabricated. It also limits the model’s generalization ability across different noise environments or pathological patterns. Moreover, supervised learning methods require clean PCG data for training, which is difficult to obtain and limits progression.
%


With these challenges, we propose a self-supervised PCG noise reduction model to address the issue of insufficient clean data. Additionally, we integrate contrastive learning to enable the model to adapt to several pathological patterns while further augmenting the data to enhance its generalization ability. This approach aims to balance the effectiveness of denoising and the preservation of pathological features, im- proving robustness in practical applications.

The main contributions of this paper are listed as follows.
\begin{itemize}
    \item
    An end-to-end self-supervised PCG denoising model and data augmentation to address the scarcity of data, improving generalization and clinical implementation.

    \item 
    Contrastive learning to combat blur and hallucination in different pathological modes in an unsupervised manner preserves pathological features.

    \item 
    An evaluation of ‘denoising’ methods and feature preservation through despite external noise, and the performance degradation of the classification model.

\end{itemize}

The remainder of this paper is outlined as follows: Section II reviews related works. Section III describes the database and methodology. Section IV presents the results and discussion. Finally, Section V concludes the paper.

%% file: sec2_related_works.tex
\section{Background and Related Work}
\subsection{Diagnosis based on PCG}
The advancement of high-quality PCG datasets and deep learning technologies has promoted PCG segmentation and classification. Datasets such as CinC2016 \cite{liu2016open} and Yaseen \cite{yaseen2018classification} have provided a baseline for evaluating classification performance. Several deep learning models for diagnosis from abnormal PCG findings have demonstrated promising performance.
Some studies have enhanced the ability of diagnostic PCG models to capture temporal features by incorporating feature pre- extraction and integrating techniques such as Long Short-Term Memory (LSTM) \cite{al2022lightweight, alkhodari2021convolutional, radha2024raw}. Abbas et al. \cite{abbas2022automatic} proposed a model based on the convolutional vision transformer to identify and classify PCG signals into five distinct categories and achieved promising classification performance on the Yaseen dataset.
%
Maity et al. \cite{maity2023transfer} successfully transferred the YAMNET (Yet Another Mobile Net) from the audio event detection task to PCG diagnosis using transfer learning.
Although they have demonstrated excellent performance, their applicability in real-world scenarios remains limited, especially in noisy environments.


A key challenge in the practice of diagnostic PCG models is noise robustness. In Maity et al.
\cite{maity2023transfer},
although the YAMNET model exhibits the ability of noise robustness, the input signal is only filtered with a 10th-order band-pass IIR Butterworth filter, and it was reported that the classification performance was significantly impacted in the noisy environment. In the hospital noise as 0 dB, the sensitivity was decreased from 99.59\% to 84.3\%.
Most of these diagnostic models are trained on ideal environments or datasets. Thus, the performance degradation of these medical diagnostic models in the real-world environment has raised concerns about their practical effectiveness.

%

%


\subsection{PCG Noise Reduction}
Numerous PCG denoising algorithms that outperform basic low-pass and band-pass filters have been proposed, enhancing the noise robustness of the classification models. Although band-pass and low-pass filters are straightforward, they are often insufficient to handle PCG’s complex time-frequency characteristics. DWT-based approaches have gained attention in PCG noise reduction because of their multi-resolution analysis features, enhancing signal suppression and reconstruction. Researchers have explored different wavelet bases,
thresholds, and decomposition levels to seek optimal performance.
$Sym16$ enhanced symmetry and orthogonality, which enable them to preserve the structural integrity of signals while minimizing phase distortion, and are applied in this field \cite{rouis2020effectiveness}. Moreover, with the characteristics of strong time-frequency localization, the $morlet$ wavelet, which combines a sinusoid with a Gaussian envelope, has also been employed for handling complex oscillatory signals effectively~\cite{mohan2020group}. 

They have shown significant effectiveness in PCG noise reduction and have become the mainstream traditional approaches in this field, however, they require complex parameter tuning and fail to capture the latent time-frequency characteristics of the PCG. Inspired by the DWT filter’s decomposition and reconstruction, some deep learning-based PCG noise reduction algorithms have been proposed and have demonstrated excellent performance. Gonzalez et al. \cite{gonzalez2023robust} 
proposed a 2D-U- Net for PCG noise reduction. This approach converts PCGs to the spectrogram, then trains U-Net to denoise the PCGs. Denoising is performed on the PCG signals with an SNR of -5\textasciitilde10 dB under additive white Gaussian noise (AWGN), additive pink Gaussian noise (APGN), speech, and real PCG background noise. The average improvement ranges from 11.85 dB to 17.60 dB. This method is the first to use U- Net for PCG denoising. However, this method relies on the Short-Time Fourier Transform (STFT) and its inverse, which affects time efficiency and introduces loss during the conversion. Ali et al. \cite{ali2023end} 
further developed this to an End-to-End approach. They also replaced U-Net’s skip connection with a Bidirectional Long Short-Term Memory (Bi-LSTM) skip connection to improve its performance for time series data. In this study, although the authors did not evaluate the model’s performance on typical colored noise, they introduced lung, hospital, and realworld noise to assess the model’s performance, reporting denoising results superior to existing methods.

\subsection{Self-supervised Noise Reduction}
Environmental influences and limitations in equipment accuracy often make obtaining clean data challenging and almost unrealistic. This issue is widespread across both image and sound datasets used in medical research. Lehtinen et al.
\cite{lehtinen2018noise2noise} proposed a promising solution for image denoising. They assumed that the noise present in multiple acquisitions of the same sample is randomly independent. It is possible to obtain clean data by fitting one noisy data sample to another noisy sample from the same source.
This method has garnered widespread attention in the image noise reduction field and quickly evolved into numerous versions, such as Noise2Void \cite{krull2019noise2void}, which performs denoising using only noisy images without relying on any clean images, Noise2Self \cite{batson2019noise2self}, which leverages the self-similarity of images for denoising learning, and ZeroShot Noise2Noise \cite{mansour2023zero}, which can denoise without any training and noise distribution.
These methods have also been adapted to speech \cite{kashyap2021speech}, medical image \cite{wu2019consensus}, and ECG denoising \cite{liu2025self}. They continue to make breakthroughs in denoising tasks where clean data is hard to obtain, particularly in image and medical data denoising, offering continuous possibilities.


%% file: sec3_database_and_method.tex
\section{Database and Method}
\subsection{Database}
In this study, the Yaseen dataset \cite{yaseen2018classification} is utilized as the primary dataset, which provides a balanced five-class classification of normal and pathological PCGs. This dataset is collected from multiple sources, sampled at 8000 Hz, and a single sample is a 3-cycle heart sound signal. There are five categories in the database, one normal category (N) and four abnormal categories, the four abnormal categories are: aortic stenosis (AS) mitral stenosis (MS) mitral regurgitation (MR) mitral valve prolapse (MVP), with a total of 1000 (200 audio files for each category). Meanwhile, we refer to the public hospital noise \cite{shams_nafisa_ali_samiul_based_shuvo_2021}, lung sound dataset \cite{FRAIWAN2021106913}, and the audiomentation library to synthesize noisy PCG data.

\subsection{Proposed Method}
\subsubsection{Improved LU-Net}
\begin{figure}[h]
    \centering
    \includegraphics[width=0.95\linewidth]{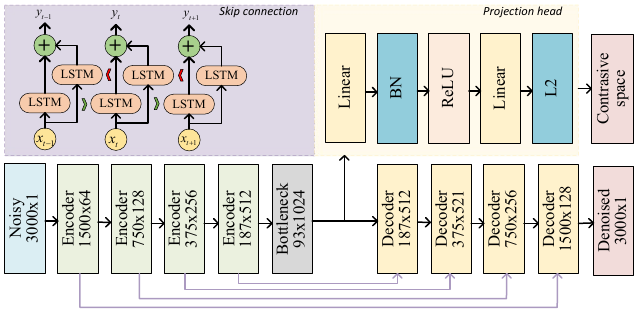}
    \caption{LU-Net with projection head.}
    \label{fig:unet}
\end{figure}

The architecture of our noise reduction model is illustrated in Fig. \ref{fig:unet}, where a four-level U-Net is employed to compress and reconstruct 1D PCG signals for noise suppression. To enhance feature retention, Bidirectional Long Short-Term Memory (BiLSTM) units are incorporated into the skip connections, allowing the decoder to recover fine-grained temporal details from the original heart sound. The depth of the U-Net is reduced, and the dropout layers are introduced to avoid overfitting. Additionally, the bottleneck representations are projected into a contrastive space $Z$ via a projection head, facilitating the analysis of latent space.

\subsubsection{Contrastive Loss}
\begin{figure}[h]
    \centering
    \includegraphics[width=0.95\linewidth]{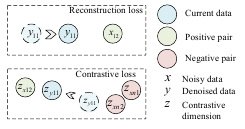}
    \caption{Contrastive learning and its loss function.}
    \label{fig:contra}
\end{figure}

The SSC-UNet aims to denoise PCGs by integrating reconstruction and contrastive learning objectives.
PCG denoising models typically use MSE as the reconstruction loss
to drive the reconstructed signal to fit the clean signal.
To address the blurring and hallucination issues of different PCGs without labels, we use contrastive loss to cluster PCGs in contrastive space $Z$. In a training batch, the same PCG samples in different distorted versions are deemed as positive pairs, all the others are negative pairs, thus constructing the contrastive loss, as shown in Fig. \ref{fig:contra}.
The total loss is formulated as a weighted sum of the reconstruction and contrastive losses, and the model parameters are updated via backpropagation.
In this way, the model can avoid confusion between different modes of PCGs.

\subsubsection{N2N Trainning Process}
Given a batch of noisy PCG signals, additional noise will be injected before feeding the perturbed signals into the SSC-UNet model.
The network parameters are initialized randomly, and a projection head is incorporated for contrastive representation learning.
During each training iteration, the model processes the distorted input to generate a denoised output, and the reconstruction loss is computed based on the mean squared error between the output and the original noisy input.
Simultaneously, bottleneck features are projected into a latent space, where a contrastive loss is computed by maximizing the similarity between different augmentations of the same sample while minimizing the similarity between different samples.
This process is repeated until convergence, yielding a trained SSC-UNet model capable of effectively reducing noise.

\begin{figure}[h]
    \label{alg:lstm_unet_denoise}
    \renewcommand{\algorithmicrequire}{\textbf{Input:}}
    \renewcommand{\algorithmicensure}{\textbf{Output:}}
    \removelatexerror
    \begin{algorithm}[H]
        \caption{SSC-UNet N2N training process}
        \begin{algorithmic}[1]
            \REQUIRE Noisy PCG $x_{1}$, noise injection $\mathcal{N}$, batch size $m$, learning rate $\alpha$, contrastive weight $\lambda$, temperature $\tau$  
            \ENSURE Denoised PCG $y_{\text{1}}$, Projection $z$, Weight $\Theta$  
            \STATE Initialize SSC-UNet parameters $\Theta$ with random weights
            \STATE Initialize projection head $g(\cdot)$ for contrastive learning
            \REPEAT
                \FOR{each batch $\{x_{\text{1}}^{(i)}\}_{i=1}^m$}
                    \STATE Add distortion: ${x}_2^{(i)} \gets n(x_{\text{1}}^{(i)}), n \sim \mathcal{N}$
                    \STATE Forward: ${y}_{\text{1}}^{(i)} \gets \text{SSC-UNet}({x_2}^{(i)})$
                    \STATE $L_{\text{recon}} \gets \frac{1}{m}\sum_{i=1}^m \| {y}_{\text{1}}^{(i)} - x_{\text{1}}^{(i)} \|^2$
                    \STATE Project bottleneck features: $z^{(i,j)} \gets g(h_{\text{bottleneck}}^{(i,j)})$
                    \FOR{each sample $i$}
                        \STATE Positive pairs: $(z^{(i,j)}, z^{(i,k)})_{j\neq k}$
                        \STATE Negative pairs: $(z^{(i,j)}, z^{(l,m)})_{l\neq i}$
                        \STATE $L_{\text{contra}} \gets -\log\frac{\exp(\text{sim}(z^{(i,j)},z^{(i,k)})/\tau)}{\sum_{n\neq i}\exp(\text{sim}(z^{(i,j)},z^{(n,m)})/\tau)}$
                    \ENDFOR
                    \STATE Total loss: $L \gets L_{\text{recon}} + \lambda L_{\text{contra}}$
                    \STATE Update $\Theta$ via backpropagation: $\Theta \gets \Theta - \alpha \nabla_\Theta L$
                \ENDFOR
            \UNTIL{Convergence}
            \RETURN Trained SSC-UNET $\Theta$
        \end{algorithmic}
    \end{algorithm}
\end{figure}

%% file: sec4_result_and_discussion.tex
\section{Result and discussion}
\subsection{Experiment Setup}
The experimental data is derived from the Yaseen dataset after preprocessing, consisting of 13,326 original segments and 232,380 segments augmented with noise or non-temporal transformations. Each segment has a sampling rate of 2,000 Hz and a duration of 1.5 seconds. The dataset is split in an 8:1:1 ratio for training, validation, and testing. Model training is conducted on an A2000 Ada GPU. The experiments are implemented using Python 3.9 and PyTorch 2.5.1, with CUDA 12.4. The training process adopts a learning rate of $0.6 \times 10^{-3}$ and a batch size of 16, with each method trained for 60 epochs.
\begin{figure}[h]
    \centering
    \includegraphics[width=0.9\linewidth]{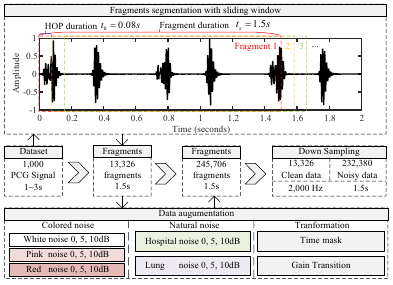}
    \caption{Preprocessing and data augmentation.}
    \label{fig:preprocessing}
\end{figure}
\subsection{Evaluation Metric}


The Signal-to-Noise Ratio (SNR) is a measure used to quantify the quality of a signal relative to background noise. A higher SNR indicates a clearer signal with less noise interference, while a lower SNR suggests that the noise level is comparable to or higher than the signal itself.
\begin{table*}[htbp]
    \centering
    \caption{Output PCG SNR with input PCG SNR under noise (dB)}
    \label{tab:pcg_snr}
    \begin{tabular}{lccccccccccccccc}
        \toprule
        \multicolumn{1}{c}{Model} & \multicolumn{3}{c}{White} & \multicolumn{3}{c}{Pink} & \multicolumn{3}{c}{Red} & \multicolumn{3}{c}{Hospital} & \multicolumn{3}{c}{Lung (unseen)} \\
        \cmidrule(lr){2-4} \cmidrule(lr){5-7} \cmidrule(lr){8-10} \cmidrule(lr){11-13} \cmidrule(lr){14-16}
        & 0 & 5 & 10 & 0 & 5 & 10 & 0 & 5 & 10 & 0 & 5 & 10 & 0 & 5 & 10 \\
        \midrule
        AE  & 11.39 & 11.95 & 12.2 & 8.81 & 11.25 & 12.87 & 10.56 & 12.98 & 14.9 & 7.22 & 10.17 & 12.2 & 8.29 & 10.64 & 13.03 \\
        U-Net  & 16.19 & 18.12 & 19.18 & 13.82 & 15.93 & 17.58 & 15.11 & 16.89 & 18.11 & 12.86 & 14.62 & 16.97 & 13.94 & 16.26 & 17.69 \\
        LU-Net & 17.4 & 19.86 & 21.35 & 14.8 & 17 & 18.8 & 16.19 & 18.05 & 19.4 & 13.8 & 16.22 & 18.38 & 16.93 & 18.89 & 19.9 \\
        \textbf{N2N}    & 17.63  & 20.42 & 22.1 & 14.5 & 17.3 & 19.24 & 16.28 & 18.37 & 19.92 & 14 & 16.62 & 19.07 & 16.99 & 19.2 & 20.52 \\
        \textbf{CN2N}   & 17.27 & 20.11 & 22.15 & 14.26 & 16.83 & 18.99 & 15.71 & 17.85 & 19.34 & 12.98 & 16.08 & 18.83 & 16.24 & 18.45 & 19.6 \\
        \bottomrule
    \end{tabular}
\end{table*}

We trained a five-class 1D-CNN+BiLSTM model for PCG classification. This model achieved over 99\% sensitivity (Se), specificity (Sp), and accuracy (Acc) in a noise-free environment. It was utilized to assess the impact of noise on PCG classification performance. The implementation of denoising methods aims not only to remove noise but also to preserve pathological features. Specifically, the evaluation examines whether the classification model can still identify these pathological markers after denoising.  




\subsection{Preprocessing and Augmentation}
The preprocessing and augmentation flow is shown in Fig. \ref{fig:preprocessing}.
We utilize the PCG recordings from the Yaseen dataset and segment them into 13,326 PCG segments of 1.5 seconds each using a hop time of 0.08 seconds. The parameters were selected from a study that found the best classification performance when tuning the preprocessing parameters for this dataset \cite{maity2023transfer}.

We apply various noise augmentation and enhancement algorithms, expanding the dataset to 245,760 enhanced segments. This augmentation includes the normal colored noise, such as white, pink, and red noise at 0, 5, 10\,\text{dB}. Noise from realistic environments is also considered, such as hospital and lung noise at 0, 5, 10\,\text{dB}. We also applied techniques such as temporal masking, gain transition, and the addition of randomly sustained noise to further enhance the dataset. 
Finally, the signals are downsampled to 2,000\,\text{Hz} to reduce data size and conserve computational resources.

\subsection{Result}

The clean segments in the dataset are used to evaluate the model's performance by calculating the output SNR when the noise injection level ranged from 0 to 10 dB, and used 500 sets of data to evaluate the average denoising performance. It is worth noting that lung noise was not included in the denoising training process. For performance comparison, we reproduced and optimized the existing denoising models shown in Table \ref{tab:pcg_snr} based on the open source code of previous studies. The improved AE, UNet, LU-Net, and our method were compared. Although the model was never exposed to completely clean PCG signals during training, our method still outperformed the existing methods in terms of signal-to-noise ratio. In addition, with 0dB open source hospital noise as input, our method achieved an SNR of 14dB, which is higher than FCN (4.21dB), U-Net (4.06dB), and LU-Net (5.99dB) mentioned in the literature \cite{ali2023end}, surpasses state of the art. Although the introduction of contrastive learning will slightly reduce SNR, it can effectively alleviate the ``hallucination" problem common in UNet-based denoising methods, that is, the model tends to generate signals that do not exist to minimize the loss.


\begin{figure}[h]
    \centering
    \includegraphics[width=1\linewidth]{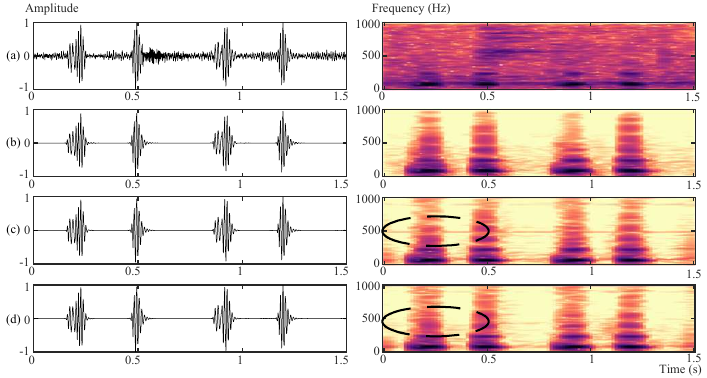}
    \caption{Waveforms and spectrograms of a normal PCG. (a) noisy signal, (b) original signal (hospital noise at 5 dB SNR), denoised signals using (c) N2N (d) CN2N, respectively.}
    \label{fig:stft}
\end{figure}


We selected a representative noisy PCG sample to assess the preservation of pathological features in PCGs after denoising and to examine potential hallucination issues. We analyzed spectrograms before and after denoising with different methods, as shown in Fig. \ref{fig:stft}. This model can reduce noise and mitigate the blurring and hallucination artifacts observed in traditional U-Net-based denoising methods with contrastive learning.

\begin{figure}[h]
    \centering
    \includegraphics[width=0.9\linewidth]{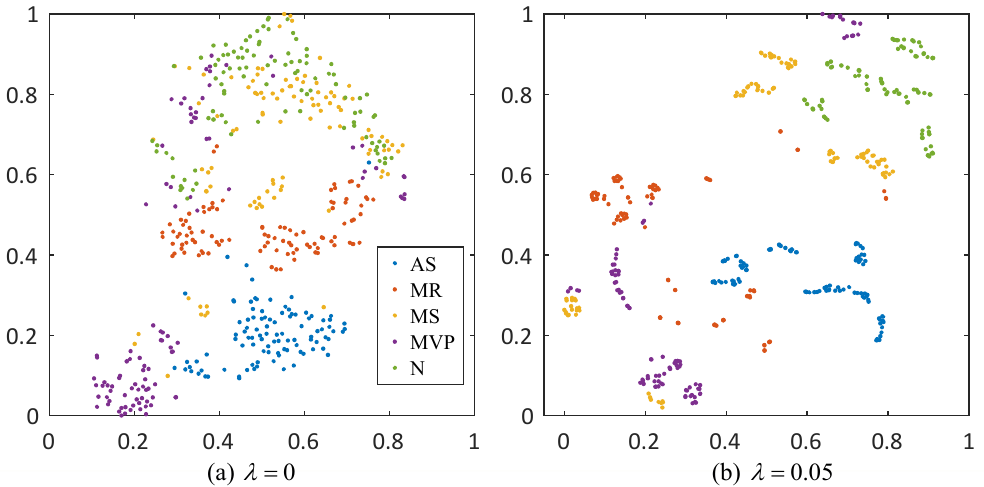}
    \caption{t-SNE of the projection head with contrastive learning.}
    \label{fig:tsne}
\end{figure}

The contrastive learning approach enables the effective separation of different PCG patterns and targeted denoising in an unsupervised manner. Additionally, we employ the t-SNE method to map the contrastive space obtained from the projection head into a two-dimensional space, where different PCG patterns are color-coded, as shown in Fig. \ref{fig:tsne}. Compared to the latent space distribution of PCGs in  Fig. \ref{fig:tsne} (a), the incorporation of contrastive learning in  Fig. \ref{fig:tsne} (b) allows the encoder to achieve the clustering and separation of different PCG patterns during encoding without labels. This reduces blurring artifacts after denoising and enhances the model’s adaptability to various pathological PCG patterns.

\begin{figure}[h]
    \centering
    \includegraphics[width=0.9\linewidth]{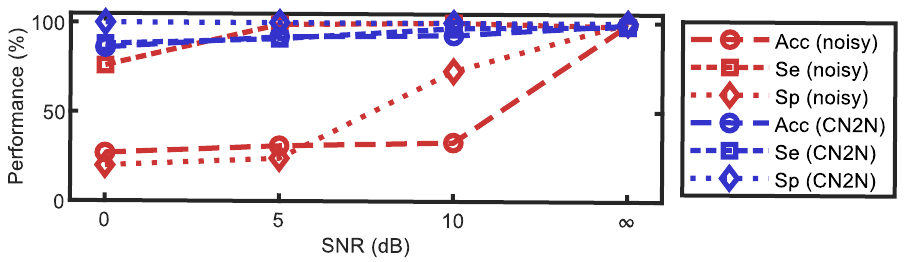}
    \caption{The performance degradation of classification models under noise and the suppression by the denoising model.}
    \label{fig:deg}
\end{figure}

To evaluate the effectiveness of our approach in preserving pathological features, we applied the denoising approach to PCGs under 10 dB hospital noise conditions and examined noise impact on the degradation of Se, Sp, and Acc classification performance, as shown in Fig. \ref{fig:deg}. Without any denoising strategy, hospital noise significantly degrades the performance of our pre-trained 1D-CNN+BiLSTM model in five-class classification tasks, namely Se, Sp, and Acc. We observed that applying denoising measures can effectively alleviate the degradation of the classification model, demonstrating this approach can improve the SNR and effectively preserve pathological features.

\subsection{Discussion}
Based on the experiments, the SSC-UNet outperforms traditional UNET methods in denoising PCG without clean segments. This holds significant implications for PCG analysis in scenarios where clean data is scarce. With contrastive learning, the model can separate pathological PCG patterns without labels, adapting to various types of PCGs and mitigating blurring and hallucination issues.

Although the LSTM skip connections enable the denoising model to capture some temporal information, it still performs poorly when dealing with non-independent and identically distributed noise. In the future, we plan to explore transformers or attention-based mechanisms for PCG denoising. The end-to-end approach has significantly reduced resource usage and improved its practical applicability. However, while neural networks provide excellent denoising performance, they also increase computational complexity, making real-time applications impractical. One of our future directions is to explore how to distill this effective denoising method into time-frequency masks to enhance its real-time capability. This real-time performance will enable edge deployment while ensuring noise reduction performance, which is expected to promote the advancement of medical IoT devices.


%% file: sec5_conclusion.tex
\section{conclusion}
This study introduces an end-to-end method for self-supervised PCG denoising based on Noise2Noise and contrastive learning to alleviate the performance degradation of CHD detection models under noise. This approach does not rely on a large amount of clean data. It can achieve denoising without seeing clean data. In addition, the contrastive learning combats the hallucination of the model and retains the PCG characteristics. Experimental results show that this method exhibits strong robustness to various types of noise, effectively suppresses background noise, and enhances the discriminability of PCG signals. In addition, this method is not limited to PCG denoising, but can also be extended to other physiological signal processing tasks, making it suitable for different application scenarios. Future research can further integrate expert knowledge, improve denoising strategies for specific frequency bands, and improve the accuracy of PCG-based CHD detection.


%% file: sec6_ack.tex
\section{Acknowledgement}
This work has emanated from research conducted with the financial support of Taighde Éireann – Research Ireland under Grant number 22/PATH-S/10763.